\documentclass{desyproc}

\begin{document}
%------------------------------------
\title{High-Intensity Probes of Axion-Like Particles}

%for single authors the superscripts are optional
\author{{\slshape Babette D\"{o}brich and Holger Gies}\\[1ex]
Theoretisch-Physikalisches Institut, Friedrich-Schiller-Universit\"{a}t Jena \\
\& Helmholtz Institute Jena, Max-Wien-Platz 1, D-07743 Jena, Germany}

% if the proceedings are available online (e.g. at Indico)
% please enter the contribution ID or file_name below for the DOI
%\contribID{32}
\contribID{dobrich\_babette}

% TO THE CONFERENCE EDITORS: 
% please update the following information      
% before sending the template to the authors
% \confID{800}  % if the conference is on Indico uncomment this line
\desyproc{DESY-PROC-2010-03}
\acronym{Patras 2010} % if you want the Acronym in the page footer uncomment this line
\doi  % if there is an online version we will register DOIs

\maketitle

\begin{abstract}
With continuously increasing intensities, modern laser systems can become a valuable tool for the search for axions and axion-like particles.
As conventional setups of axion searches cannot easily accommodate the usage of a high-intensity laser system, we propose a novel, purely laser-based setup in which the occurrence 
of a frequency shift is an observable for the axion-photon interaction. 

\end{abstract}

\section{Motivation}
Optical probes of the axion-like particle (ALP) parameter space, such as polarimetric measurements  %\cite{Maiani:1986md} 
or ''Light-shining-through-walls'' setups, %\cite{Sikivie:1983ip}
are enhanced by applying strong external (electro-)magnetic fields $B$ over a large spatial extent $L$. 
Conventionally, dipole magnets are used to modify the propagation of probe beams. But with the remarkable increase in 
available laser intensity over the past years, multi-terawatt laser systems have become competitive with dipole magnets 
in providing $B\times L$ and can be expected to eventually exceed them. Still, the limited temporal and spatial extent of 
pulsed high-intensity beams disfavors standard setups used for ALP search and requires different 
observables which are particularly useful for purely laser-based experiments. 
Such an observable can, e.g., be a diffraction pattern \cite{Tommasini:2010fb}, or a frequency shift \cite{Dobrich:2010hi,Homma:2010jc}. 
The latter will be the subject of this contribution.

\section{Photon-Axion dynamics}

As we are interested in the effects of the nonlinear interaction of laser
photons due to the presence of an axion or ALP, we start from the equations of motion for the two
fields:
\begin{eqnarray}
\partial_{\mu}\partial^{\mu}\phi+m^2\phi-\frac{1}{4}g F_{\mu\nu}\tilde{F}^{\mu\nu} &=& 0 \label{eq:axion_EOM} \\
\partial_{\mu}F^{\mu\nu}-g(\partial_{\mu}\phi)\tilde{F}^{\mu\nu} &=& 0 \label{eq:photon_EOM}\ .
\end{eqnarray}
For the following discussion, we split the field strength tensor $F_{\mu\nu}$ into contributions of a probe field $F_{\mu\nu}^{\mathrm{in}}$ and two external 
high-intensity beams $F_{\mu\nu}^{\mathrm{ext}}$,
and linearize in the probe field.
Coupling to the first external beam in Eq. \eqref{eq:axion_EOM}, the probe photons can be converted into ALPs, denoted by $\phi$.
Successively, by means of Eq. \eqref{eq:photon_EOM}, the ALPs can be reconverted into photons through the second external field.
For simplicity, the following discussion will be limited to a one-dimensional setup where the probe photons propagate along the positive $z$ axis. 
Then, the above equations of motion can be solved using a Green's function approach \cite{Dobrich:2010hi,Adler:2008gk}.

In order to motivate the findings for a purely laser-based setup we first review the probe photon dynamics for static or slowly varying external 
fields spanning a 
length $L$ as e.g. provided by a dipole magnet. In this case, employing an incoming plane wave probe beam of frequency $\omega_{\mathrm{in}}$ in 
Eq. \eqref{eq:axion_EOM} yields 
via Eq. \eqref{eq:photon_EOM} an outgoing wave that carries the original frequency $\omega_{\mathrm{in}}$ after the intermediate 
propagation as an ALP, necessitating 
the use of a light-blocking wall or polarimetric measurements. 
In addition, the outgoing wave's amplitude picks up two factors of $\sin (\Delta k \frac{L}{2})/\Delta k$ with 
$\Delta k =-\omega_{\mathrm{in}}+\sqrt{\omega_{\mathrm{in}}^2-m^2}$ 
arising in the conversion and back-conversion processes, respectively, causing it to be maximal at $\Delta k \simeq 0$. 
From this it follows that the ALP search with dipole magnets 
is most sensitive for small ALP masses, cf. also Fig. \ref{fig:bounds}.

From another perspective, the sensitivity for small axion masses being maximal is related to momentum conservation 
as reflected by the requirement $\Delta k \simeq 0$ 
for both the photon-axion conversion and back-conversion process.

\section{Photon-Axion conversion in high-intensity laser fields}

\begin{wrapfigure}{r}{0.6\textwidth}
  \begin{center}
    \includegraphics[width=0.58\textwidth]{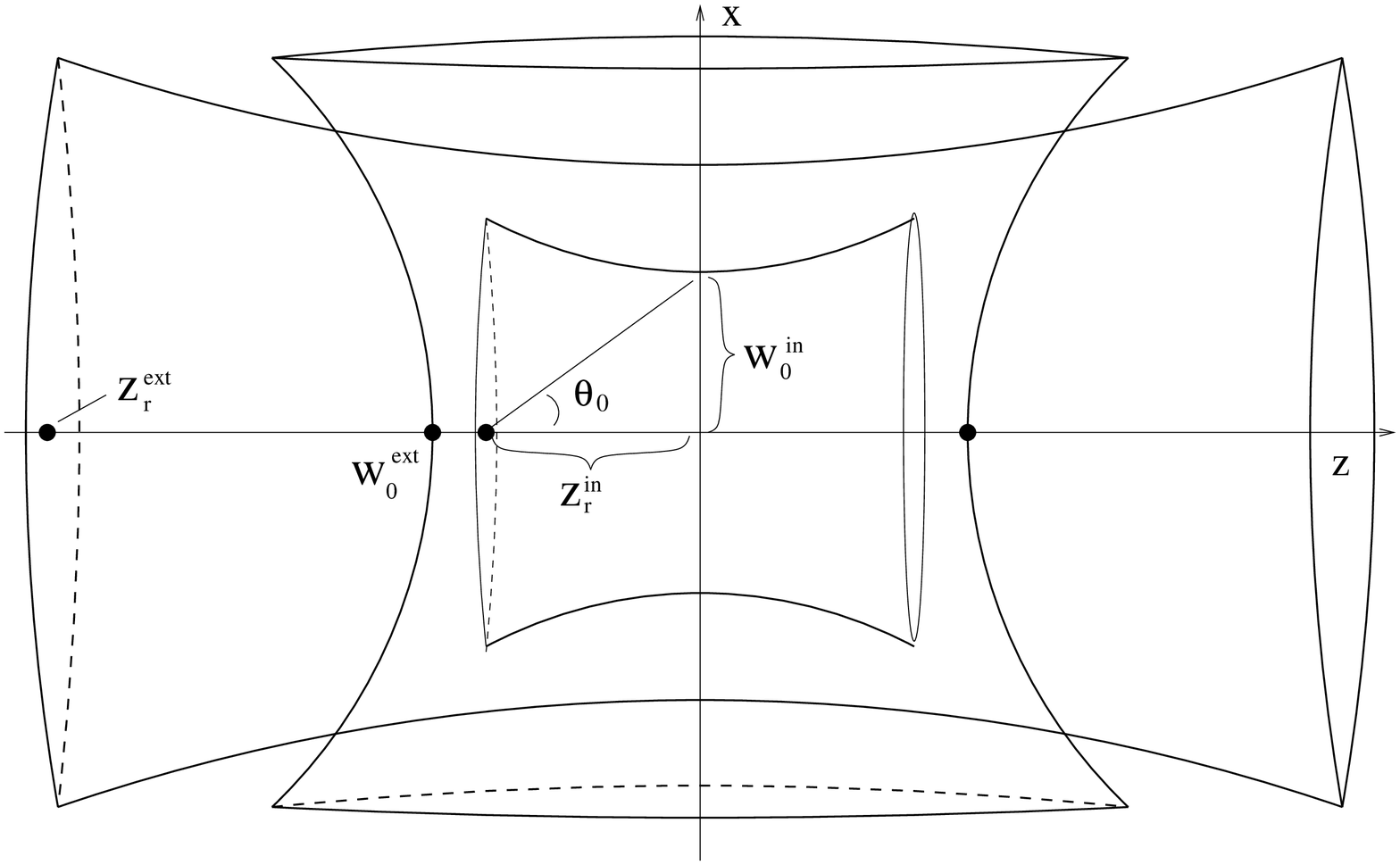}
  \end{center}
\caption{Overlap of Gaussian laser beams. The innermost probe beam 
%($\omega_{\mathrm{in}}$) 
is
embedded into two external beams propagating orthogonally 
%($\omega_{\mathrm{\bot}}$) 
and transversally 
%($\omega_{\mathrm{\parallel}}$)
to it, respectively. The waist size $w_0^{\mathrm{ext}}$ as well as the Rayleigh lengths
$z_r^{\mathrm{in}}$ and  $z_r^{\mathrm{ext}}$ determine the extent of the beam foci.}

\label{fig:setup}
\end{wrapfigure}

From the preceding consideration it becomes obvious that employing lasers as external fields will modify the momentum 
balance as they carry a further frequency scale. This changes the sensitivity characteristics with 
respect to the ALP mass. Moreover, in accordance with energy conservation, the frequency of the outgoing photon will in general be modified. 
This happens in close analogy to the processes of sum-frequency generation (SFG) and difference-frequency generation (DFG) known from nonlinear optics. 
The frequency of the outgoing electromagnetic wave will be the sum or difference of the incident frequencies and, analogous to the phase matching 
conditions within SFG and DFG, the wave vectors of the external beams (being different in general) enter the 
requirements\footnote{Introducing high-intensity lasers as external fields changes also the functional form of the conversion amplitude since one 
has to employ a Gaussian beam form \cite{Davis:1979}. Also, the length scale $L$ will be substituted by the natural spatial extent of Gaussian beams, 
being the waist size $w_0$ and the Rayleigh length $z_r$ of the beams.} $\Delta k \simeq 0$.
In summary, we thus suggest the measurement of a frequency shift $\omega_{\mathrm{out}}\neq\omega_{\mathrm{in}}$ of probe beam photons as an 
observable of photon-ALP interaction, which is feasible for unsuppressed interaction amplitudes, i.e. $\Delta k \simeq 0$ for the conversion 
and back-conversion process. This perfectly accommodates the features of the high-intensity beam making the use of a light-blocking wall or polarimetric measurements superfluous.
With hindsight, we choose a configuration in which the photon-axion conversion is mediated by an external beam of frequency $\omega_{\bot}$ 
propagating orthogonally to the $z$ axis, while the back-conversion process from the ALP to a photon is due to a counter-propagating beam with $\omega_{\parallel}$, cf. Fig. \ref{fig:setup}.

Then, as within SFG and DFG, one obtains from Eq. \eqref{eq:axion_EOM} ALP partial waves carrying frequencies $\omega_{\mathrm{ax}}^{\pm}=\omega_{\mathrm{in}}\pm\omega_{\bot}$, 
while the corresponding `phase matching condition' becomes 
\begin{equation}
\Delta k_{\bot}^{\pm}= -\omega_{\mathrm{in}}+\delta_{\mathrm{T}}\sqrt{(\omega_{\mathrm{in}}\pm\omega_{\bot})^2-m^2} \stackrel{!}{\simeq} 0 \ , \label{eq:k_j_plus} 
\end{equation}
with $\delta_{\mathrm{T}}=\pm 1$ in case of transmission and reflection of the ALP wave, respectively.

Above, $\Delta k_{\bot}^{+}= 0$ is realized in the case of transmission (i.e. $\delta_{\mathrm{T}}=+1$) 
for $m= m_{\parallel} \equiv \sqrt{\omega_{\bot}^2+2\omega_{\mathrm{in}}\omega_{\bot}}$, whereas $\Delta k_{\bot}^{-}= 0$ 
cannot satisfy the requirement of a non-negative ALP frequency $\omega_{\mathrm{ax}}$.
Thus, only the `SFG solution' $\omega_{\mathrm{ax}}^{+}=\omega_{\mathrm{in}}+\omega_{\bot}$ is kept in the following.

Note that, since the wave vector of the external $\bot$ beam has no component along $z$, it does not enter \textit{explicitly} in the 
momentum conservation in Eq. \eqref{eq:k_j_plus}. This is different for the back-conversion process mediated by the $\parallel$ beam. 
The outgoing electromagnetic partial waves now carry frequencies $\omega_{\mathrm{out}}=\omega_{\mathrm{in}}+\omega_{\mathrm{\bot}}\pm\omega_{\mathrm{\parallel}} $,
whilst momentum conservation requires that
\begin{equation}
\Delta k_\parallel^{\pm}= - \sqrt{(\omega_{\mathrm{in}}+\omega_\bot)^2-m^2} +
\delta_{\mathrm{T}} (\omega_{\mathrm{in}}+\omega_\bot \pm \omega_\parallel)\pm\omega_\parallel \stackrel{!}{\simeq} 0 \label{eq:delta_k_pl_vac} \ .
\end{equation}
As argued above, Eqs. \eqref{eq:k_j_plus} and \eqref{eq:delta_k_pl_vac} have to hold simultaneously for a feasible sensitivity while we demand 
$\omega_{\mathrm{out}}\neq \omega_{\mathrm{in}}$ as an observable for the photon-ALP conversion.
Requiring positivity of all frequencies, it can be checked that this is only possible for DFG and transmission of the wave, i.e. $\delta_{\mathrm{T}}
=+1$ in Eq. \eqref{eq:delta_k_pl_vac}.
In particular, choosing $\omega_\bot=2 \omega_\parallel$, we see that $\Delta k_\parallel^{-}=\Delta k_\bot^{+}=0$ for axion masses $m=m_{\parallel}$ 
yielding $\omega_{\mathrm{out}}=\omega_{\mathrm{in}}+\omega_{\mathrm{\parallel}}$. This constitutes an observable of the ALP-photon 
interaction\footnote{In the discussed setting, the photon-ALP conversion can also be induced by the ($\bot$) oriented field and back-conversion can be 
due to the counter-propagating ($\parallel$) field as the interaction order is not assessable for synchronized pulses. The conversion process with the opposite order 
$\bot \leftrightarrow \parallel$ results in
$\omega_{\mathrm{out}}=\omega_{\mathrm{in}}-\omega_\parallel$ for axion masses around $m_\parallel=2\sqrt{\omega_{\mathrm{in}}\omega_\parallel}$
and $\omega_\bot=2 \omega_\parallel$, thus defining a second 'resonant mass' besides $m_\bot$ \cite{Dobrich:2010hi}.}.
As the resonant masses $m_{\parallel}$ and $m_{\bot}$ are of the same order of magnitude as the laser frequency scales being $ \mathcal{O}(\mathrm{eV})$, purely 
laser-based searches are \textit{complementary} to standard dipole setups.
Of course, to facilitate the detection of the frequency shift requires $\omega_{\mathrm{out}}$ to lie feasibly outside the spectral widths $\Delta \omega$ of all 
interacting beams.

At last, it is worth emphasizing that the condition $\omega_\bot=2 \omega_\parallel$ is in fact an enormous experimental advantage since second harmonic 
generation is a standard technique even for high-intensity lasers, thus requiring the employment of only \textit{one} external high-intensity beam.

\section{Discovery potential}

\begin{wrapfigure}{r}{0.7\textwidth}
  \begin{center}
\vspace{-1.3cm}
\includegraphics[scale=0.55]{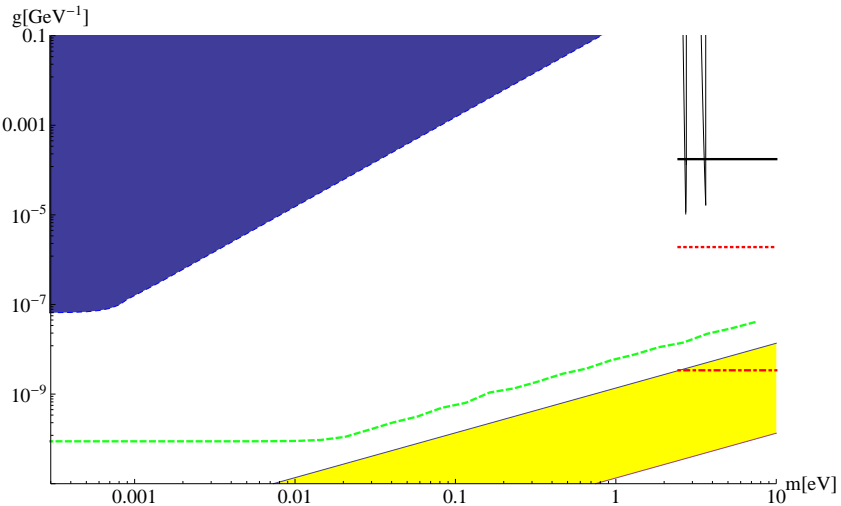}
\end{center}
\caption{ALP exclusion bounds for
  ALPS \cite{Ehret:2010mh} (blue-shaded area), and CAST \cite{CAST} (green-dashed line) in comparison to a setup
  involving the laser systems available in Jena \cite{IOQ}, denoted by black wedges.  The black line indicates the principle exclusion
  bounds at this setup employing a conventional OPA system, while the red-dotted line gives an analogous estimate for ELI \cite{ELI}. The red dot-dashed
  line suggests the requirements at ELI for entering the parameter
  regime of typical QCD axion models which are given by a yellow band.}
\label{fig:bounds}
\end{wrapfigure}

An estimate of the ALP discovery potential for purely laser-based searches is given in Fig. \ref{fig:bounds}, see \cite{Dobrich:2010hi} for details.
Here, the black wedge-like curves and the black line correspond to a feasible setup at IOQ \cite{IOQ} employing the multi-TW class laser JETI
% ($\omega_{\mathrm{in}}=1.55 \mathrm{eV}$) 
as probe and the PW class laser
POLARIS as external beam,
%($\omega_{\parallel}=1.2 \mathrm{eV}$) 
which could be realized in the near future.  For the estimate, single-photon detection is assumed for the frequency-shifted 
photons\footnote{For higher photon statistics the number of shots $N_{\mathrm{shot}}$ can be increased ($\mathcal{O}(100)$ per day at IOQ \cite{IOQ}).}.
The wedge-like structures of the bounds around $m_{\parallel}$ and $m_{\bot}$ are a consequence of the fixed frequencies within the setup.

Using optical parametric amplification (OPA) for the probe beam, a larger range of the mass-coupling plane can be explored, as indicated 
by the black line.
However, as OPA limits intensity, the sensitivity to the ALP coupling is decreased.
Nevertheless, we see that a setup at IOQ could provide the strongest model-independent \cite{Jaeckel:2006xm} bounds for ALP masses at 
$\mathcal{O}(\mathrm{eV})$.

Even smaller ALP coupling values can potentially be probed at the future exawatt class facility ELI \cite{ELI}.
Already with single shot measurements it can be possible to almost complement the currently best laboratory
bounds provided by ALPS \cite{Ehret:2010mh} in the $\mathcal{O}(\mathrm{eV})$ mass range as shown by the red dotted line in Fig. \ref{fig:bounds}.
In order to surpass even the CAST \cite{CAST} bounds on solar axions, one would require $N_{\mathrm{shot}}N_{\mathrm{in}}\approx 10^{26}$ at ELI 
(dash-dotted line). 
This constitutes a rather ambitious aim, but would allow for a direct probe of the QCD axion parameter space 
(given as a yellow band in Fig. \ref{fig:bounds}), making it a worthwhile task for the future.  

\section{Conclusions}

The rapid increase in available laser intensity strongly suggests to investigate the potential of high-intensity lasers for axion and ALP search. 
As argued, a possible observable in this context is the measurement of a probe beam frequency shift. This is particularly useful since the limited temporal 
and spatial extent of the pulses disfavors conventional setups such as polarimetric measurements and Light-shining-through-walls.
In summary, high-intensity probes of ALPs 
%could complement the bounds of established experiments in the $\mathcal{O}(\mathrm{eV})$ mass range and 
can constitute a 
new tool in the general quest \cite{Jaeckel:2010ni} for weakly interacting slim particles.
B.D. thanks the organizers of the 6th Patras Workshop for the opportunity to present this work. Support by the DFG through grants SFB/ TR18, GRK1523, and Gi328/5-1 is
gratefully acknowledged.

%\section*{Acknowledgments}
%\vspace{+1cm}

%

% ****************************************************************************
% BIBLIOGRAPHY AREA
% ****************************************************************************

\begin{footnotesize}

\end{footnotesize}

% ****************************************************************************
% END OF BIBLIOGRAPHY AREA
% ****************************************************************************


\begin{thebibliography}{99}



%\cite{Tommasini:2010fb}
\bibitem{Tommasini:2010fb}
  D.~Tommasini and H.~Michinel,
  %``Light by light diffraction in vacuum,''
  Phys.Rev.A {\bf 82}, 011803 (2010),
%(2010);
  [arXiv:1003.5932 [hep-ph]].
  %%CITATION = PHRVA,A82,011803;%%



%\cite{Dobrich:2010hi}
\bibitem{Dobrich:2010hi}
  B.~Dobrich and H.~Gies,
  %``Axion-like-particle search with high-intensity lasers,''
  JHEP {\bf 1010}, 022 (2010),
 [arXiv:1006.5579 [hep-ph]].
  %%CITATION = JHEPA,1010,022;%%

%\cite{Homma:2010jc}
\bibitem{Homma:2010jc}
  K.~Homma {\it et al.},
 % D.~Habs and T.~Tajima,
  %``Probing semi-macroscopic vacua by high fields of lasers,''
  arXiv:1006.4533 [quant-ph].
  %%CITATION = ARXIV:1006.4533;%%

  
 \bibitem{Adler:2008gk}
 S.~L.~Adler {\it et al.}, 
 %J.~Gamboa, F.~Mendez and J.~Lopez-Sarrion,
  %``Axions and 'Light Shining Through a Wall': A Detailed Theoretical
  %Analysis,''
  Annals Phys.\  {\bf 323}, 2851 (2008).
  %[arXiv:0801.4739 [hep-ph]].
  %%CITATION = APNYA,323,2851;%%
  
\bibitem{Davis:1979}
L.W.~Davis,
%{\sl Theory of electromagnetic beams},
Phys.\ Rev.\ A {\bf 19}, 1177-1179 (1979).

%\bibitem{shg}
%  P.~A.~Franken, A.~E.~Hill, C.~W.~Peters and G.~Weinreich
%  %``Generation of optical harmonics,''
%  Phys.\ Rev.\ Lett.\  {\bf 7}, 118-119 (1961).
%  %%CITATION = PRLTA,88,095005;%%

%\cite{Jaeckel:2006xm}
\bibitem{Jaeckel:2006xm}
  J.~Jaeckel {\it et al.}, 
%E.~Masso, J.~Redondo, A.~Ringwald and F.~Takahashi,
  %``The Need for Purely Laboratory-Based Axion-Like Particle Searches,''
  Phys.\ Rev.\  D {\bf 75}, 013004 (2007),
 [arXiv:hep-ph/0610203].
  %%CITATION = PHRVA,D75,013004;%%
 
\bibitem{ELI}
http://www.extreme-light-\\
infrastructure.eu/

%\bibitem{POLARIS}
%M.~Hornung et. al
% Temporal pulse control of a multi-10 TW diode-pumped Yb:Glass laser 
%Appl. Phys B (2010)
%doi: 10.1007/s00340-010-3952-7.

\bibitem{IOQ}
http://www.physik.uni-\\
jena.de/inst/ioq//start- \\
Engl.html

%\cite{Ehret:2010mh}
\bibitem{Ehret:2010mh}
  K.~Ehret {\it et al.} ,
  %``New ALPS Results on Hidden-Sector Lightweights,''
  Phys.\ Lett.\  B {\bf 689}, 149 (2010),
  [arXiv:1004.1313 [hep-ex]].
  %%CITATION = PHLTA,B689,149;%%
  
  \bibitem{CAST}
%\cite{Arik:2008mq}
%\bibitem{Arik:2008mq}
  E.~Arik {\it et al.} ,
  % [CAST Collaboration],
  %``Probing eV-scale axions with CAST,''
  JCAP {\bf 0902}, 008 (2009),
  [arXiv:0810.4482].
  %%CITATION = JCAPA,0902,008;%%

%\cite{Jaeckel:2010ni}
\bibitem{Jaeckel:2010ni}
  J.~Jaeckel and A.~Ringwald,
  %``The Low-Energy Frontier of Particle Physics,''
  arXiv:1002.0329 [hep-ph].
  %%CITATION = ARXIV:1002.0329;%%



\end{thebibliography}
\end{document}